\documentclass[epj]{svjour}

\usepackage{amsmath}
\usepackage{graphicx}
\usepackage{ifpdf}
\usepackage[numbers,sort&compress]{natbib}
\usepackage{rotating}
\usepackage{url}
\usepackage{xspace}

\ifpdf
	\usepackage{pdfsync}
	\usepackage{epstopdf}
\else
\fi

\usepackage[breaklinks]{hyperref}

\newcommand{\apriori}{\textit{a priori}\xspace}
\newcommand{\ie}{i.e.}
 
\newcommand{\eg}{e.g.}
\newcommand{\GLO}{GLO\xspace}

\newcommand{\foreign}{\textit}
\newcommand{\defineterm}{\emph}

\newcommand{\largeparens}[1]{\left ( #1 \right )}

\newcommand{\largecurly}[1]{\left \{ #1 \right \}}

\newcommand{\set}{\largecurly}

\newcommand{\floor}[1]{\left \lfloor #1 \right \rfloor}

\newcommand{\mathpuncspace}{\quad}
\newcommand{\mathcomma}{\mathpuncspace,}
\newcommand{\mathperiod}{\mathpuncspace.}

\newcommand{\matelem}[3]{#1_{#2#3}}

\newcommand{\numedges}{\ensuremath{m}}
\newcommand{\numverts}{\ensuremath{n}}
\newcommand{\adjelem}[2]{\matelem{A}{#1}{#2}}
\newcommand{\degree}[1]{\ensuremath{k_{#1}}}
\newcommand{\edge}[2]{\ensuremath{(#1, #2)}}
\newcommand{\vertex}{\ensuremath}

\newcommand{\modularity}{\ensuremath{Q}}
\newcommand{\modularitychange}[2]{\Delta\modularity_{#1#2}}
\newcommand{\modularitynorm}{\frac{1}{2\numedges}}

\newcommand{\maxmodchange}[1]{\Delta\modularity_{#1}^{\mathrm{max}}}
\newcommand{\maxmod}[1]{\textbf{#1}}

\newcommand{\clustwt}[1]{\ensuremath{W_{#1}}}
\newcommand{\clustvol}[1]{\ensuremath{K_{#1}}}
\newcommand{\clustmodnorm}{\lambda}
\newcommand{\interclustwt}[2]{\ensuremath{w_{#1#2}}}

\newcommand{\worstcase}[1]{\ensuremath{O\largeparens{#1}}}
\newcommand{\dendroheight}{\ensuremath{h}}
\newcommand{\numpasses}{\ensuremath{p}}

\newcommand{\schemex}{\ensuremath{X}}
\newcommand{\schemey}{\ensuremath{Y}}

\newcommand{\probability}[1]{\ensuremath{P\largeparens{#1}}}
\newcommand{\mutinfosymbol}{\ensuremath{I}}
\newcommand{\mutinfo}[2]{\mutinfosymbol\largeparens{#1, #2}}
\newcommand{\normalizedmutinfosymbol}{\ensuremath{I_\mathrm{norm}}}
\newcommand{\normalizedmutinfo}[2]{\normalizedmutinfosymbol\largeparens{#1, #2}}
\newcommand{\entropysymbol}{\ensuremath{H}}
\newcommand{\entropy}[1]{\entropysymbol\largeparens{#1}}
\newcommand{\relentsum}[3]{\sum_{#3} #1 \log \frac{#1}{#2}}
\newcommand{\entropysum}[2]{-\sum_{#2} #1 \log #1}

\newcommand{\sizeratio}{\ensuremath{R}}
\newcommand{\commsize}[1]{\ensuremath{n_{#1}}}

\graphicspath{{pdffigs/}{epsfigs/}{mpsfigs/}}

\begin{document}

\title{
Detecting hierarchical and overlapping network communities using locally optimal modularity changes
}
\titlerunning{Network communities using locally optimal modularity changes}

\author{Michael J. Barber}
\institute{AIT Austrian Institute of Technology GmbH, Foresight and Policy Development Department, Vienna, Austria
\email{michael.barber@ait.ac.at}
}
\authorrunning{Barber}

\date{\today}

\abstract{
Agglomerative clustering is a well established strategy
for identifying communities in networks. Communities are successively
merged into larger communities, coarsening a network of actors into
a more manageable network of communities. The order in which merges
should occur is not in general clear, necessitating heuristics for
selecting pairs of communities to merge.  We describe a hierarchical clustering 
algorithm based on a local optimality property. For each edge in the network, we associate the
modularity change for merging the communities it links. For each
community vertex, we call the preferred edge that edge for which
the modularity change is maximal. When an edge is preferred by both
vertices that it links, it appears to be the optimal choice from
the local viewpoint. We use the locally optimal edges to define
the algorithm: simultaneously
merge all pairs of communities that are connected by locally optimal
edges that would increase the modularity, redetermining the locally
optimal edges after each step and continuing so long as the modularity
can be further increased. We apply the algorithm to model and
empirical networks, demonstrating that it can efficiently produce
high-quality community solutions. 
We relate the performance and implementation details to the structure of the resulting community hierarchies. 
We additionally consider a complementary local clustering algorithm, describing how to identify overlapping communities based on the local optimality condition.
}

\maketitle

\section{Introduction}\label{sec:intro}

A prominent theme in the investigation of networks is the identification
of their community structure. Informally stated, network
communities are subnetworks whose constituent vertices are strongly
affiliated to other community members and comparatively weakly affiliated
with vertices outside the community; several formalizations of this
concept have been explored (for useful reviews, see Refs.~\cite{PorOnnMuc:2009,For:2010}). The
strong internal connections of community members is often accompanied by
greater homogeneity of the members, e.g., communities in the World Wide
Web as sets of topically related web pages or communities in scientific
collaboration networks as scientists working in similar research
areas. Identification of the network communities thus can facilitate
qualitative and quantitative investigation of relevant subnetworks whose
properties may differ from the aggregate properties of the network as
a whole.

Agglomerative clustering is a well established strategy 
for identifying a hierarchy of communities in networks. Communities are successively
merged into larger communities, coarsening a network of actors into
a more manageable network of communities. The order in which merges
should occur is not in general clear, necessitating heuristics for
selecting pairs of communities to merge.  

A key approach to community identification in networks is from \citet{New:2004a},
who used a greedy agglomerative clustering algorithm to search for
communities with high modularity \citep{NewGir:2004}. In this algorithm,
pairs of communities are successively merged based on a global optimality
condition, so that the modularity increases as much as possible with each
merge. The pairwise merging ultimately produces a community hierarchy that
is structured as a binary tree. The structure
of the hierarchy closely relates to both the quality of  the solution
and the efficiency of its calculation; modularity is favored by uniform
community sizes \citep{DanDiaAre:2006,BarCla:2009} while rapid computation
is favored by shorter trees \citep{ClaNewMoo:2004}, so both are
favored when the community hierarchy has a well-balanced binary tree structure,
where the sub-trees at any node are similar in size.
But the greedy algorithm may produce unbalanced community hierarchies---the
hierarchy may even be dominated by a single large community that absorbs
single vertices one-by-one \citep{WakTsu:2007}, 
causing the hierarchy to be unbalanced at all levels. 

In this work, we propose a new agglomerative clustering strategy
for identifying community hierarchies in networks. We replace the global optimality
condition for the greedy algorithm with a local optimality condition. The
global optimality condition holds for communities \( c \) and \( c^\prime \) 
when no other pair of communities could be merged so as to increase the
modularity more than would merging \( c \) and \( c^\prime \). The local
optimality condition weakens the global condition, holding when no pair of
communities, one of which is either \( c \) or \( c^\prime \), could be
merged to increase the modularity more than would merging \( c \) and 
\( c^\prime \). The essentials of the clustering strategy follow directly:
concurrently merge communities that satisfy the local optimality condition
so as to increase the modularity, re-establishing the local optimality conditions and
repeating until no further modularity
increase is possible. The concurrent formation of communities encourages
development of a cluster hierarchy with properties favorable both to
rapid computation and to the quality of the resulting community solutions.

\section{Agglomerative clustering}\label{sec:background}

\subsection{Greedy algorithms}

Agglomerative clustering \citep{HasTibFri:2009,JaiMurFly:1999} is an 
approach long used \citep{War:1963} for classifying
data into a useful hierarchy. The approach is based on assigning the
individual observations of the data to clusters, which are fused or
merged, pairwise, into successively larger clusters. The merging process
is frequently illustrated with a dendrogram, a tree diagram showing the
hierarchical relationships between clusters; in this work, we will also
refer to the binary tree defined by the merging process as a dendrogram,
regardless of whether it is actually drawn.

Specific clustering algorithms depend on defining a measure of the
similarity of a pair of clusters, with different measures
corresponding to different concepts of clusters. Additionally, a rule
must be provided for selecting which merges to make based on their
similarity. Commonly, merges are selected with a greedy strategy, where
the single best merge is made and the similarity recalculated
for the new cluster configuration, making successive merges until only
a single cluster remains. The greedy heuristic will not generally identify the 
optimal configuration, but can often find a good one.

\subsection{Modularity}

Agglomerative clustering has seen much recent use for investigating
the community structure of complex networks 
(for a survey of agglomerative clustering and other community identification approaches, see Refs.~\citep{For:2010,PorOnnMuc:2009}).
The dominant approaches follow \citet{New:2004a} in searching for
communities (\ie, clusters) with high modularity \(\modularity\). 
Modularity assesses community strength for a partition of the 
\(\numverts\) network vertices into disjoint sets, and is defined \citep{NewGir:2004} as
\begin{equation}
	\modularity = \modularitynorm \sum_{c} \sum_{i,j \in c} \largeparens{\adjelem{i}{j} - \frac{ \degree{i} \degree{j}}{2\numedges} }
	\mathcomma
	\label{eq:modularity}
\end{equation}
where 
the \(\adjelem{i}{j}\) are elements of the adjacency matrix for the graph, 
\(\numedges\) is the number of edges in the graph, 
and \(\degree{i}\) is the degree of vertex \(i\), \ie, \(\degree{i} = \sum_{j} \adjelem{i}{j} \). 
The outer sum is over all clusters \(c\), 
the inner over all pairs of vertices \((i,j)\) within \(c\). 

With some modest manipulation, Eq.\nobreakspace \textup {(\ref {eq:modularity})} can be 
written in terms of cluster-level properties
and in a form suitable as well for use with weighted graphs:
\begin{equation}
	\modularity = \clustmodnorm \sum_{c} \largeparens{ \clustwt{c} - \clustmodnorm \clustvol{c}^2 }
	\mathcomma
	\label{eq:clustmod}
\end{equation}
where
\begin{align}
	\clustwt{c} & = \sum_{i,j \in c} \adjelem{i}{j} \\
	\clustvol{c} & = \sum_{i \in c} \degree{i} \\
	\clustmodnorm & = \largeparens{ \sum_{c} \clustvol{c} }^{-1} \label{eq:clustmodnorm}
	\mathperiod
\end{align}
Here, \( \clustwt{c} \) is a weight of edges internal to cluster \(c\),
measuring the self-affinity of the cluster constituents; \( \clustvol{c} \) 
is a form of volume for cluster \(c\), analogous to the graph volume; 
and \(\clustmodnorm\) is a scaling factor equal to
\(1/2\numedges\) for an unweighted graph. 
Other choices for \(\clustmodnorm\) may also be suitable \citep{BarCla:2009}, 
but we will not consider them further.

Edges between vertices in different clusters \(c\) and \(c^\prime\) may
also be described at the cluster level; denote this edge by \edge{c}{c^\prime}. Edge \edge{c}{c^\prime} 
has a corresponding symmetric inter-cluster weight \(\interclustwt{c}{c^\prime}\), defined by
\begin{equation}
	\interclustwt{c}{c^\prime} = \sum_{i \in c} \sum_{j \in c^\prime} \adjelem{i}{j}
	\label{eq:interclustwt}
	\mathperiod
\end{equation}
Using \interclustwt{c}{c^\prime}, we can describe the merge process entirely in terms of cluster properties. When two clusters \(u\) and \(v\) are merged into a new cluster \(x\), it will have 
\begin{align}
	\clustwt{x} & = \clustwt{u} + \clustwt{v} + 2\interclustwt{u}{v} \\
	\clustvol{x} & = \clustvol{u} + \clustvol{v}
	\mathperiod
\end{align}
The inter-cluster weights for the new cluster \(x\) will be
\begin{equation}
	\interclustwt{x}{y} = \interclustwt{u}{y} + \interclustwt{v}{y}
\end{equation}
for each existing cluster \(y\), excluding \(u\) and \(v\). 
The modularity change \(\modularitychange{u}{v}\) is
\begin{equation}
	\modularitychange{u}{v} = 2 \clustmodnorm \largeparens{ \interclustwt{u}{v} - \clustmodnorm \clustvol{u}\clustvol{v}    }
	\mathperiod
	\label{eq:modularitychange}
\end{equation}
From Eq.\nobreakspace \textup {(\ref {eq:modularitychange})}, it is clear that modularity can only increase when \( \interclustwt{u}{v} > 0 \) and, thus, when there are edges between vertices in \(u\) and \(v\). 

With the above, we can view a partition of the vertices as an equivalent graph of clusters or communities; merging two clusters equates to edge contraction. The cluster graph is readily constructed from a network of interest by mapping the original vertices to vertices representing singleton clusters and edges between the vertices to edges between the corresponding clusters. For a cluster \(c\) derived from a vertex \(i\), we initialize \( \clustwt{c} = 0 \) and \( \clustvol{c} = \degree{i} \).

\subsection{Modularity-based greedy algorithms}

\citet{New:2004a} applied a greedy 
algorithm to finding a high modularity partition of network vertices 
by taking the similarity measure to be the 
change in modularity \(\modularitychange{u}{v}\). In this approach, 
\(\modularitychange{u}{v}\) is evaluated for each inter-cluster edge, and 
a linked pair of clusters leading to maximal increase in modularity is selected for the merge. 
A naive implementation of this greedy algorithm constructs 
the community hierarchy and identifies the level  in it with greatest modularity
in worst-case time \(\worstcase{(\numedges + \numverts)\numverts}\), 
where \( \numedges \) and \( \numverts \) are, respectively, the numbers of edges and 
vertices in the network.

Finding a partition giving the global maximum in \(\modularity\) is
a formally hard, NP-complete problem, equivalent to finding the ground
state of an infinite-range spin glass \citep{ReiBor:2006}. We should thus
expect the greedy approach only to identify a high modularity partition in
a reasonable amount of time, rather than to provide us with the global
maximum. Variations on the basic greedy algorithm may be developed
focusing on increasing the community quality, reducing the time taken,
or both.

Likely the most prominent such variation is the implementation described
by \citet{ClaNewMoo:2004}. While neither the greedy strategy nor
the modularity similarity measure is altered, the possible merges
are tracked with a priority queue implemented using a binary heap, allowing rapid determination
of the best choice at each step. This results in a worst-case
time of \(\worstcase{\numedges\dendroheight\log\numverts}\), where
\(\dendroheight\) is the height of the resulting dendrogram. Thus, the
re-implementation is beneficial when, as for many empirical networks of interest,
the dendrogram is short, ideally
forming a balanced binary tree with height equal to 
\(\floor{\log_2 \numverts}\), where \(\floor{x}\) denotes the integer part of \( x \). 
But the dendrogram need not be short---it may be
a degenerate tree of height \(\numverts\), formed when all singleton
clusters are merged one-by-one into the same cluster. Such a dendrogram
results in \(\worstcase{\numedges\numverts\log\numverts}\) time, 
worse than for the naive implementation.

Numerous variations on the use of the change in modularity
have been proposed for use with greedy algorithms, with
some explicitly intended to provide a shorter, better balanced
dendrogram. We note two in particular. First, \citet{DanDiaAre:2006}
consider the impact that heterogeneity in community size has on the
performance of clustering algorithms, proposing an altered modularity
as the similarity measure for greedy agglomerative clustering. Second,
\citet{WakTsu:2007} report encountering poor scaling behavior for the
algorithm of \citeauthor{ClaNewMoo:2004}, caused by merging communities
in an unbalanced manner; they too propose several modifications to the
modularity to encourage more well-balanced dendrograms. In both papers,
the authors report an improvement in the (unmodified) modularity
found, even though they were no longer directly using modularity to
select merges---promoting short, well-balanced dendrograms can promote better
performance both in terms of time taken and in the quality of the
resulting communities.

Alternatively, the strategy by which merges are selected may be changed,
while keeping the modularity as the similarity measure, giving rise to
the multistep greedy (MSG) algorithm \citep{SchCaf:2008,SchCaf:2008a}. In
the MSG approach, multiple merges are made at each step, instead of just
the single merge with greatest increase in the modularity. The potential
merges are sorted by the change in modularity \(\modularitychange{u}{v}\) they produce; merges are
made in descending order of \(\modularitychange{u}{v}\), so long as (1) the
merge will increase modularity and (2) neither cluster to be merged has
already been selected for a merge with greater \(\modularitychange{u}{v}\). The
MSG algorithm promotes building several communities concurrently, avoiding
early formation of a few large communities. Again, this leads to shorter,
better balanced dendrograms with improved performance both in terms of
time and community quality.

More drastic changes to the basic greedy hierarchical clustering scheme are
also possible. \Citet{BloGuiLamLef:2008} describe a two-phase algorithm
consisting of first identitying a local optimum in the modularity by
repeatedly reassigning individual vertices to the communities where they
make the maximal contribution to modularity, and second constructing
a new, weighted network where those communities are the vertices;
these two phases are repeated until no further modularity increases
are possible. Based on application to sample networks, the algorithm is
reported to rapidly determine high-modularity community solutions, but
the resulting hierarchy is generally no longer a binary tree. Effectively,
the algorithm avoids formation of the unbalanced portions of the hierarchy
by allowing multiple vertices to be merged into a community in one step instead
of successively, thereby compressing uninformative portions of the hierarchy
while retaining fewer, hopefully relevant details.

When required for clarity, we will refer to the original greedy strategy as single-step greedy (SSG). 
Additionally, we will  restrict our attention to an implementation following \citet{ClaNewMoo:2004}.

\section{Clustering with local optimality}\label{sec:locopt}

\subsection{Local optimality}

Greedy community detection algorithms are intrinsically global algorithms,
drawing upon information from across the entire network to select
which communities to merge. Yet it is instructive to consider what can
be said about the merges on a local scale, and, in particular, about
the globally optimal merge selected in the SSG algorithm. Globally, if clusters \(u\)
and \(v\) are selected for the merge, 
then \( \modularitychange{u}{v} \geq \modularitychange{x}{y} \) for all clusters \(x\) and \(y\).
Restricting to the edges incident on the clusters, this indicates
\begin{align}
	\modularitychange{u}{v} & \geq \modularitychange{u}{y} \qquad \forall y \label{eq:preferredv} \\
	\modularitychange{u}{v} & \geq \modularitychange{x}{v} \qquad \forall x \label{eq:preferredu}
	\mathperiod
\end{align}
Informally, the two clusters each have the other as the best choice of merge; 
these local properties are necessary, but not sufficient, conditions for the 
global optimality condition in SSG---an edge may satisfy the local conditions, but
not cause the greatest modularity change in the network. 

\begin{figure}
	\includegraphics[width=\columnwidth]{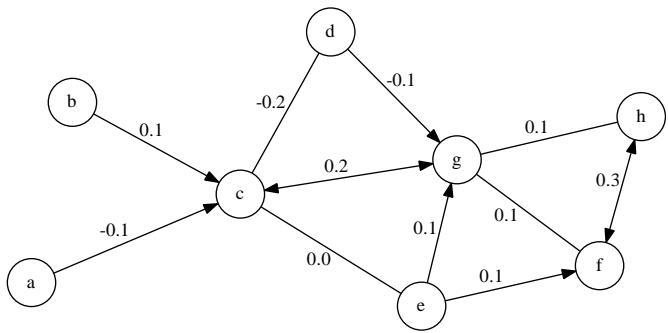}
	\caption{Preferred and locally optimal edges. Each edge is labeled with its modularity change \(\modularitychange{u}{v}\), which is the basis for determing the merge preferences shown with arrows. Edges with a single arrowhead are preferred edges for the vertex at the tail of the arrow, but not for the vertex at the head of the arrow. Edges with arrowheads at each end are preferred for both vertices; these are locally optimal edges. Those edges without arrowheads are not preferred by either of the linked vertices.  }
	\label{fig:locoptnet}
\end{figure}

We may use this idea to classify the inter-cluster edges. Call
an edge \edge{u}{v} \defineterm{preferred} for vertex \vertex{u} or \vertex{v}
if Eq.\nobreakspace \textup {(\ref {eq:preferredv})} or Eq.\nobreakspace \textup {(\ref {eq:preferredu})} holds, respectively;
similarly refer to the corresponding merge as preferred. If the edge 
is preferred for both \vertex{u} and \vertex{v}, call it \defineterm{locally optimal.}
We illustrate preferred and locally optimal edges in Fig.\nobreakspace \ref {fig:locoptnet}.

\subsection{Greedy clustering using local optimality} \label{sec:locoptgreedy}

We can base an agglomerative clustering algorithm on merging along
the locally optimal edges in the network, determining whether any edges
become locally optimal as a consequence, and repeating this until no
locally optimal edges remain. With such an approach, we discourage
the formation of unbalanced dendrograms by allowing multiple merges to
occur concurrently, thus favoring shorter dendrograms 
and---given a suitable implementation---more efficient computation. 
The approach lies somewhere between SSG and MSG clustering, 
featuring concurrent formation of clusters like MSG, but selecting 
merges with a generalization of the condition in SSG.

For the most part, it is straightforward to define a precise algorithm
from this idea. One complication is the presence of vertices with multiple
locally optimal edges incident upon them. These edges can lead to, for example,
a state where edges \edge{u}{v} and \edge{u}{w} are locally optimal, 
but \edge{v}{w} is not locally optimal. Thus, if we make both locally optimal
merges, we produce a combined cluster of \(\set{u, v, w}\) which also
includes the locally suboptimal merge. But to exclude merging \(v\)
and \(w\), we must then only make one of the locally optimal merges.
In this work, we adopt the latter approach, arbitrarily selecting one of the
locally optimal merges. 

The resulting algorithm is:
\begin{enumerate}
\item For each edge \edge{u}{v}, evaluate \(\modularitychange{u}{v}\).
\item For each vertex \(v\), identify the maximum modularity change \(\maxmodchange{v}\) from all incident edges.
\item For each edge \edge{u}{v}, determine if it is locally optimal by testing \( \modularitychange{u}{v} = \maxmodchange{u} = \maxmodchange{v} \). If, in addition, \( \modularitychange{u}{v} > 0 \), edge \edge{u}{v} is a candidate merge.
\item If there are no candidate merges, stop. Otherwise, for each candidate, merge the corresponding clusters, so long as neither cluster has so far been changed in the current iteration.
\item Begin a new iteration from step 1.
\end{enumerate}
The order of iteration in step 4 will have an effect on the resulting
community hierarchy when vertices have multiple locally optimal edges. In the
implementation used in this work, we iterate through the edges in an
arbitrary order that is uncorrelated with the modularity changes 
\(\modularitychange{u}{v}\). As the algorithm greedily selects edges based on local optimality, 
we call it \GLO clustering---greedy, local optimality clustering.

When the \GLO algorithm terminates, no remaining edge will support a
positive change in modularity; otherwise, one or more edges \edge{u}{v}
would have \( \modularitychange{u}{v} > 0 \), and thus there would
be at least one candidate merge---that edge with the greatest 
\( \modularitychange{u}{v} \). The clusters at termination have greater
modularity than at any earlier iteration in the algorithm, since merges
are only made when they increase the modularity.

Note that the \GLO algorithm generally terminates only having formed
the sub-trees of the dendrogram for each cluster rather than the
full dendrogram with single root. If the full dendrogram is needed,
additional cluster merges can be made by using an alternate greedy
algorithm. Here, we follow the above steps for \GLO clustering, but
drop the requirement that \( \modularitychange{u}{v} > 0 \)---all locally 
optimal edges become candidate merges. This laxer
condition is always satisfied by at least at least the edge with greatest
\( \modularitychange{u}{v} \), so the merge process continues until all
edges have been eliminated and only a single cluster remains.

Implementing the \GLO algorithm presents no special difficulties. The
needed properties of the clusters (\(\clustwt{v}\), \(\clustvol{v}\),
and \(\interclustwt{u}{v}\)) can be handled as vertex and edge attributes
of a graph data structure. Straightforward implementation of the above
steps can be done simply by iterating through the \(\numedges\)
edges, leading to \(\worstcase{\numedges}\) worst-case time complexity
for each of the \(\numpasses\) iterations of the merge process,
or \(\worstcase{\numedges\numpasses}\) overall worst-case time complexity. A
simple optimization of this basic implementation strategy is to keep
track of the \( \maxmodchange{v}\) values and a list of corresponding
preferred edges, recalculating these only when merges could lead to
changes; this does not change the worst case time complexity from
\(\worstcase{\numedges\numpasses}\), but does notably improve
the execution speed in practice.

The above estimates of time complexity have the shortcoming that they
are given not just in terms of the size of the network, but also in terms 
of an outcome of the algorithm---the number of iterations
\(\numpasses\). There is no
clear \apriori relation between \(\numpasses\) and the network
size, but we may place bounds on \(\numpasses\). First, the algorithm
merges at least one pair of clusters in each iteration, so \(\numpasses\)
is bounded above by \(\numverts\). Second, the algorithm involves any
cluster in at most one merge in an iteration, so \(\numpasses\) must
be at least the height \(\dendroheight\) of the dendrogram. This gives
\begin{equation}
	\numverts > \numpasses \geq \dendroheight \geq \floor{\log_2 \numverts}
	\mathperiod
\end{equation}
Runtime of the algorithm is thus seen to be dependent on the structure of the
cluster hierarchy found, with better performance requiring a well-balanced
dendrogram. We do have reason to be optimistic that 
\(\numpasses\) will be relatively small in this case: a well-balanced
dendrogram results when multiple clusters are constructed concurrently,
which also requires fewer iterations of the algorithm.

\subsection{Local clustering using local optimality}\label{sec:clusterexpansion}

Although all merging decisions in \GLO clustering are made using
only local information, the algorithm is nonetheless a global
algorithm---the clusters possible at one point in the graph are influenced
by merges concurrently made elsewhere in the network. Yet we may
specify a local clustering algorithm: starting from a single vertex,
successively merge along any modularity-increasing, locally optimal edges incident upon it,
stopping only when no such locally optimal edges remain. In this fashion,
the modularity---an assessment of a partition of the vertices---may be
used to identify overlapping communities. 

This local algorithm functions by absorbing vertices one-by-one into a
single cluster. Unfortunately, this is exactly the behavior corresponding
to the worst case behavior for the SSG and \GLO algorithms, producing
a degenerate binary tree as the dendrogram whose height is one less than the
number of vertices in the community and conceivably is  one less than the number of
vertices in the graph. The expected time complexity is thus quadratic
in the resulting community size. Worse still, characterizing all local
clusters for the graph may require a sizable fraction of the vertices
to be so investigated, giving a worst-case time complexity that is cubic
in the number of vertices of the graph. Such an approach is thus suited
for networks of only the most modest size.

A compromise approach is possible using a hybrid of the agglomerative
and local approaches. First, determine an initial set of clusters
using the \GLO algorithm. Second, for each community, expand it using
local clustering, treating all other vertices as belonging to distinct
singleton clusters. The hybrid algorithm is still quite slow 
(and leaves the worst-case time complexity unchanged), but fast
enough to provide some insight into the overlapping community structure
of networks with tens of thousands of vertices.

\section{Results}\label{sec:results}

\subsection{Model networks}\label{sec:modelnets}

To begin, we confirm that the \GLO clustering algorithm 
is able to identify network communities by 
applying it to randomly generated
graphs with known community structure. We make use of the model graphs
proposed and implemented by \citet{LanForRad:2008}. We generate 1000
random graphs using the default parameter settings, so that each random
graph has 1000 vertices with an average degree of 15.

In Table\nobreakspace \ref {tbl:modelnets}, we show some characteristics of the results of
clustering algorithm, comparing the results to those for SSG and MSG
clustering. For the model networks, \GLO produces community solutions
that have a greater number of communities, on average, than either SSG
or MSG. The average modularity is greatest with SSG, with \GLO second
and MSG lowest. 
Modularity values are sufficiently high 
to indicate that \GLO clustering is able to recognize 
the presence of communities in the model networks.

While modularity characterizes clustering, it does not directly measure
the accuracy of the clusters. We instead assess accuracy using the
normalized mutual information \normalizedmutinfosymbol. For the joint
probability distribution \probability{\schemex, \schemey} over random
variables \(\schemex\) and \(\schemey\), 
\( \normalizedmutinfo{\schemex}{\schemey} \) is
\begin{equation}
	 \normalizedmutinfo{\schemex}{\schemey} = \frac
			{2 \mutinfo{\schemex}{\schemey}}
			{\entropy{\schemex} + \entropy{\schemey}}
	\mathcomma
	\label{eq:normmutinfo}
\end{equation}
where the mutual information \( \mutinfo{\schemex}{\schemey} \) 
and entropies \( \entropy{\schemex} \) and \( \entropy{\schemey} \) are defined
\begin{align}
	\mutinfo{\schemex}{\schemey} & = \relentsum{\probability{\schemex, \schemey}}{\probability{\schemex}\probability{\schemey}}{x,y} \label{eq:mutinfo} \\
	\entropy{\schemex} & = \entropysum{\probability{\schemex}}{x} \\
	\entropy{\schemey} & = \entropysum{\probability{\schemey}}{y} \label{eq:entropyy}
	\mathperiod
\end{align}
In Eqs.\nobreakspace  \textup {(\ref {eq:normmutinfo})} to\nobreakspace  \textup {(\ref {eq:entropyy})} , we use the typical abbreviations 
\( \probability{\schemex=x, \schemey=y} = \probability{\schemex, \schemey} \), 
\( \probability{\schemex=x} = \probability{\schemex} \), and 
\( \probability{\schemey=y} = \probability{\schemey} \).
The base of the logarithms in Eqs.\nobreakspace  \textup {(\ref {eq:mutinfo})} to\nobreakspace  \textup {(\ref {eq:entropyy})}  is arbitrary, as the computed measures only appear in the ratio in Eq.\nobreakspace \textup {(\ref {eq:normmutinfo})}.

To assess clustering algorithms with 
\( \normalizedmutinfo{\schemex}{\schemey} \), we treat the actual community
membership for a vertex as a realization of a random variable \(\schemex\)
and the community membership algorithmically assigned to the vertex as a
realization of a second random variable \schemey. The joint probability
\probability{\schemex, \schemey} is then defined by the distribution
of paired community membership over all vertices in the graph. We can
then evaluate \( \normalizedmutinfo{\schemex}{\schemey} \), finding a
result that parallels the modularity: SSG on average obtains the greatest
normalized mutual information, with \GLO second and MSG the lowest.
The high value for \( \normalizedmutinfo{\schemex}{\schemey} \)  
indicates that \GLO clustering assigns most vertices to the correct communities.

As \GLO clustering attempts to improve performance by favoring
well-balanced dendrograms, we also assess the balance of the dendrograms
using their height. Since a dendrogram is a binary tree, the optimal
height for a graph with \( \numverts \) vertices is just the integer part of 
\( \log_2 \numverts \); the extent to which the dendrogram height exceeds this value
is then indicative of performance shortcomings of the algorithm. The
random graphs considered in this section have 1000 vertices, and
therefore the optimal height is 9. The results are essentially what
one would expect: SSG, which does not attempt to favor merges leading
to balanced dendrograms, produces the tallest dendrograms on average;
MSG, which aggressively makes concurrent merges, produces the shortest
dendrograms; and \GLO, which makes concurrent merges more selectively
than MSG, produces dendrograms with heights on average between those resulting
from SSG and MSG.

\begin{table}
\centering
\caption{
Algorithm performance with model networks. Values are computed by averaging over clustering results from 1000 realizations of the random graphs proposed by \citet{LanForRad:2008}, with default parameter settings. Results shown are for the highest modularity clusters in the generated hierarchies, with the number of clusters in the partition, the corresponding modularity \( \modularity \), the normalized mutual information \( \normalizedmutinfosymbol \) comparing the algorithm output to the known community assignments, and the height \( \dendroheight \) of the dendrogram (optimal height would be 9). Uncertainties for the final significant digits are shown parenthetically. All values in each column differ significantly (\(p < 0.001 \)).
}
\label{tbl:modelnets}
\begin{tabular}{lrrrr}
\hline\hline
Algorithm & \multicolumn{1}{c}{Clusters} & \multicolumn{1}{c}{\modularity} & \multicolumn{1}{c}{\normalizedmutinfosymbol} & \multicolumn{1}{c}{\dendroheight}\\
\hline
SSG & 16.16(5) & 0.7155(2) & 0.8481(6) & 124.1(5)\\
\GLO & 25.55(5) & 0.6904(2) & 0.8379(5) & 38.3(1)\\
MSG & 15.70(5) & 0.5673(5) & 0.6457(8) & 11.94(2)\\
\hline\hline
\end{tabular}
\end{table}

\subsection{Empirical networks}

Based on the model networks considered in 
the preceding section, it appears that SSG produces the best
community solutions of the three clustering algorithms considered. But
we are ultimately not interested in  model networks---it
is in the application to real networks that we are concerned. In this
section, we consider algorithm performance with several commonly used
empirical networks. 

\begin{table}
\centering
\caption{Empirical networks under consideration. The number of vertices \(\numverts\) and edges \(\numedges\) in each network are shown.}
\label{tbl:netprops}
\begin{tabular}{lrr}
\hline\hline
Network & \multicolumn{1}{c}{\(n\)} & \multicolumn{1}{c}{\(m\)}\\
\hline
Karate club & 34 & 78\\
Dolphins & 62 & 159\\
Les Mis\'erables & 77 & 254\\
Political books & 105 & 441\\
Word adjacency & 112 & 425\\
Football & 115 & 615\\
Jazz & 198 & 2742\\
\textit{C. elegans} neural & 297 & 2148\\
Network science & 379 & 914\\
\textit{C. elegans} metabolic & 453 & 2040\\
Email & 1133 & 5452\\
Political blogs & 1222 & 16717\\
Power grid & 4941 & 6594\\
hep-th & 5835 & 13815\\
PGP users & 10680 & 24316\\
cond-mat 1999 & 13861 & 44619\\
astro-ph & 14845 & 119652\\
Internet & 22963 & 48436\\
cond-mat 2003 & 27519 & 116181\\
cond-mat 2005 & 36458 & 171735\\
\hline\hline
\end{tabular}
\end{table}

The networks considered are 
a network of friendships between members of a university karate club \citep{Zac:1977}; 
a network of frequent associations between dolphins living near Doubtful Sound, New Zealand \citep{LusSchBoiHaaSloDaw:2003}; 
a network of character co-appearances  in the novel \foreign{Les Mis\'erables} \citep{Knu:1994};
a network of related purchases of political books during the 2004 U.S. presidential election \citep{Kre:2004};
a network of word adjacency in the novel \textit{David Copperfield} \citep{New:2006a};
a network of American college football games played during the Fall 2000 season\citep{GirNew:2002};
a network of collaborations between jazz musicians \citep{GleDan:2003}; 
a network of the neural connections in the \textit{C. elegans} nematode worm \citep{WatStr:1998};
a network of co-authorships for scientific papers concerning networks \citep{New:2006a}; 
a network of metabolic processes in the \textit{C. elegans} nematode worm \citep{JeoTomAlbOltBar:2000};
a network of university e-mail interactions \citep{GuiDanDiaGirAre:2003};
a network of links between political blogs during the 2004 U.S. presidential election\citep{AdaGla:2005};
a network of the western U.S. power grid \citep{WatStr:1998};
a network of co-authorships for scientific preprints posted to the high-energy theory archive (hep-th) \citep{New:2001b};
a network of cryptographic keys shared among PGP users \citep{BogPasDiaAre:2004};
a network of co-authorships for scientific preprints posted to the astrophysics archive (astro-ph) \citep{New:2001b};
a network of the structure of the internet, at the level of autonomous systems \citep{New:2011b};
and three networks of co-authorships for scientific preprints posted to the condensed matter archive (cond-mat), based on submissions beginning in 1995 and continuing through 1999, 2003, and 2005 \citep{New:2001b}.
Several
networks feature weighted or directed edges; we ignore these, treating
all networks as unweighted, undirected simple graphs. Not all of the networks are
connected; we consider only the largest connected component from each network.
The networks vary considerably in size, with the
number of vertices \( \numverts \) and number of edges \( \numedges \)
spanning several orders of magnitude (Table\nobreakspace \ref {tbl:netprops}).

We apply SSG, MSG, and \GLO clustering algorithms to each of the empirical
networks. In Table\nobreakspace \ref {tbl:performance}, we show properties of the clusterings
produced by each of the algorithms. The properties of the community solutions differ notably
from those for the random model networks. The number of clusters produced by \GLO
clustering no longer exceeds those for SSG and MSG clustering. Instead,
the three algorithms produce similar numbers of clusters for the smaller
networks, with the SSG algorithm yielding solutions with the greatest
number of clusters for the largest networks. As well, the \GLO algorithm
tends to produce the greatest modularity values, exceeding the other
approaches for 15 of the 20 empirical networks considered, including all
of the larger networks.

The dendrograms produced for the empirical networks parallel those for
the random networks. The dendrograms resulting from the SSG algorithm
are the tallest, those from the \GLO algorithm are second, and those
from MSG the shortest. The SSG algorithm often produces 
dendrograms far taller than the ideal for a graph with a given number 
\( \numverts \) of vertices.

\begin{table*}
\centering
\caption{Comparative performance of agglomerative clustering algorithms. For each network and each algorithm, shown are the number of clusters found, the modularity \( \modularity \), and the dendrogram height \( \dendroheight \). Additionally shown for \( \dendroheight \) is the minimum height for a dendrogram for the network. }
\label{tbl:performance}
\begin{tabular}{l|rrr|ccc|rrrr}
\hline\hline
Network & \multicolumn{3}{c}{Clusters} & \multicolumn{3}{c}{\modularity} & \multicolumn{4}{c}{\dendroheight}\\
 & SSG & \GLO & MSG & SSG & \GLO & MSG & SSG & \GLO & MSG & min\\
\hline
Karate club & 3 & 4 & 4 & 0.381 & \maxmod{0.387} & 0.381 & 9 & 10 & 8 & 6\\
Dolphins & 4 & 3 & 4 & \maxmod{0.495} & 0.491 & 0.492 & 18 & 10 & 7 & 6\\
Les Mis\'erables & 5 & 6 & 6 & 0.501 & \maxmod{ 0.556 } & 0.536 & 21 & 13 & 11 & 7\\
Political books & 4 & 5 & 4 & 0.502 & \maxmod{ 0.524 } & 0.506 & 48 & 18 & 8 & 7\\
Word adjacency & 7 & 7 & 8 & \maxmod{ 0.295 } & 0.289 & 0.252 & 23 & 13 & 8 & 7\\
Football & 7 & 8 & 5 & \maxmod{ 0.577 } & 0.564 & 0.487 & 27 & 14 & 8 & 7\\
Jazz & 4 & 4 & 4 & \maxmod{ 0.439 } & 0.424 & 0.363 & 65 & 33 & 10 & 8\\
\textit{C. elegans} neural & 5 & 6 & 5 & 0.372 & \maxmod{ 0.388 } & 0.333 & 110 & 35 & 17 & 9\\
Network science & 19 & 18 & 16 & 0.838 & \maxmod{ 0.843 } & 0.836 & 47 & 18 & 13 & 9\\
\textit{C. elegans} metabolic & 11 & 10 & 9 & 0.404 & \maxmod{ 0.428 } & 0.400 & 121 & 43 & 13 & 9\\
Email & 14 & 11 & 10 & 0.510 & \maxmod{ 0.553 } & 0.487 & 333 & 60 & 16 & 11\\
Political blogs & 11 & 7 & 10 & \maxmod{ 0.427 } & 0.420 & 0.406 & 631 & 316 & 77 & 11\\
Power grid & 40 & 41 & 39 & 0.934 & \maxmod{ 0.935 } & 0.930 & 79 & 35 & 27 & 13\\
hep-th & 76 & 56 & 51 & 0.791 & \maxmod{ 0.815 } & 0.794 & 816 & 82 & 28 & 13\\
PGP users & 176 & 120 & 95 & 0.855 & \maxmod{ 0.874 } & 0.860 & 904 & 181 & 139 & 14\\
cond-mat 1999 & 165 & 77 & 71 & 0.764 & \maxmod{ 0.827 } & 0.801 & 2005 & 115 & 40 & 14\\
astro-ph & 138 & 51 & 38 & 0.622 & \maxmod{ 0.708 } & 0.642 & 3576 & 279 & 60 & 14\\
Internet & 43 & 32 & 28 & 0.630 & \maxmod{ 0.653 } & 0.644 & 3517 & 1635 & 1209 & 15\\
cond-mat 2003 & 316 & 81 & 67 & 0.671 & \maxmod{ 0.740 } & 0.690 & 5893 & 297 & 90 & 15\\
cond-mat 2005 & 472 & 77 & 70 & 0.646 & \maxmod{ 0.704 } & 0.645 & 6857 & 570 & 119 & 16\\
\hline\hline
\end{tabular}
\end{table*}

\begin{figure}
	\includegraphics[width=\columnwidth]{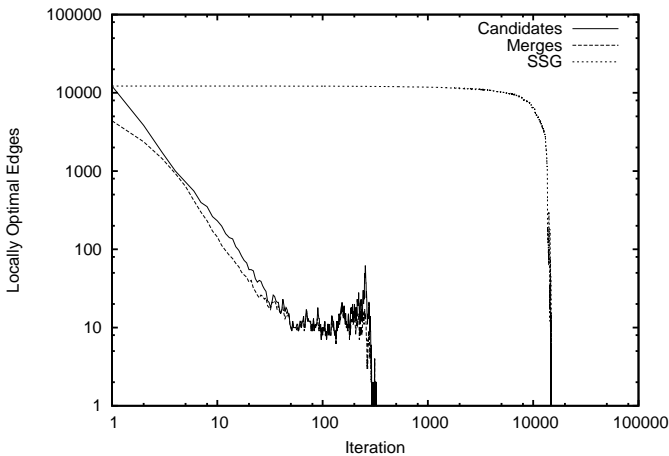}
	\caption{Number of locally optimal edges. For the astro-ph network, we show the number of locally optimal edges that are candidate merges at each iteration of the algorithm. As the algorithm is not always able to merge all the candidates, also shown are the actual number of merges made at each iteration. For comparison, we also show, for the SSG algorithm, the number of locally optimal edges that would be candidate merges in \GLO clustering.  }
	\label{fig:numlomerges}
\end{figure}

The differences between the dendrograms suggests the abundant presence
of locally optimal edges in the empirical networks. We verify this by
counting the number of candidate merges in the network for each
iteration of the \GLO and SSG algorithms. In Fig.\nobreakspace \ref {fig:numlomerges},
we show the number of candidate merges for the astro-ph network;
the other empirical networks show similar trends.

We additionally applied the local clustering scheme described in
section\nobreakspace \ref {sec:clusterexpansion}, expanding the clusters found for the
empirical network. In each case, some or all of the clusters are expanded
(Table\nobreakspace \ref {tbl:hybrid}), leading to overlapping communities. As a measure
of the degree of cluster expansion, we define a size ratio \( \sizeratio \) 
as
\begin{equation}
	\sizeratio = \frac{1}{\numverts}\sum_c \commsize{c}
	\mathcomma
\end{equation}
where \( \commsize{c} \) is the number of vertices in the expanded cluster
\( c \). The size ratio equals the expected number of clusters in which a vertex is found. 
Values of \( \sizeratio \) for the empirical networks are given
in the final column of Table\nobreakspace \ref {tbl:hybrid}. 

\begin{figure}
	\includegraphics[width=\columnwidth]{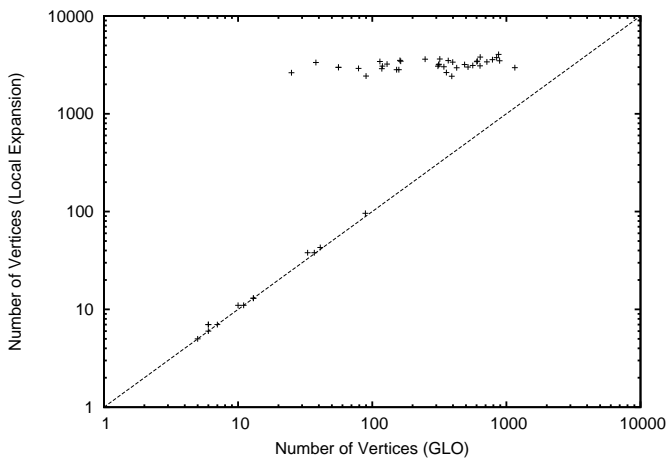}
	\caption{Expansion of communities with hybrid algorithm, consisting of the \GLO clustering algorithm followed by expansion using the local clustering algorithm. Each point corresponds to a single cluster, with the location showing the number of vertices in the cluster as determined in the \GLO stage and after the local clustering stage. The line shown indicates no expansion; all points necessarily lie on or above the line.  }
	\label{fig:expansion}
\end{figure}

The clusters do not expand uniformly.  
We illustrate this in Fig.\nobreakspace \ref {fig:expansion} using the astro-ph network. 
In this representative example, numerous clusters expand only minimally or not at all, 
while others increase in size dramatically.

\begin{table}
\centering
\caption{Cluster expansion using hybrid algorithm, consisting of the \GLO clustering algorithm followed by expansion using the local clustering algorithm. Shown are the number of clusters found in the \GLO stage, the number of those clusters that increase in size in the local clustering stage, and the size ratio \( \sizeratio \) showing an average expansion. }
\label{tbl:hybrid}
\begin{tabular}{lrrr}
\hline\hline
Network & Clusters & Expanded & \(\sizeratio\)\\
\hline
Karate club & 4 & 1 & 1.18\\
Dolphins & 3 & 1 & 1.02\\
Les Mis\'erables & 6 & 4 & 1.38\\
Political books & 5 & 5 & 1.97\\
Word adjacency & 7 & 6 & 1.31\\
Football & 8 & 7 & 1.43\\
Jazz & 4 & 4 & 1.51\\
\textit{C. elegans} neural & 6 & 6 & 2.33\\
Network science & 18 & 11 & 1.33\\
\textit{C. elegans} metabolic & 10 & 9 & 1.96\\
Email & 11 & 11 & 2.64\\
Political blogs & 7 & 3 & 1.01\\
Power grid & 41 & 23 & 1.01\\
hep-th & 56 & 45 & 4.48\\
PGP users & 120 & 43 & 2.37\\
cond-mat 1999 & 77 & 61 & 6.83\\
astro-ph & 51 & 45 & 8.20\\
Internet & 32 & 22 & 2.45\\
cond-mat 2003 & 81 & 63 & 9.39\\
cond-mat 2005 & 77 & 58 & 8.31\\
\hline\hline
\end{tabular}
\end{table}

\section{Conclusion}\label{sec:conclusion}

We have described a new agglomerative hierarchical clustering strategy
for detecting high-modularity community partitions in networks; we call 
this \GLO clustering, for greedy, local optimality clustering. At
the core of the approach is a locally optimality criterion, where
merging two communities \( c \) and \( c^\prime \) is locally optimal
when no better merge is available to either \( c \) or \( c^\prime
\). The cluster hierarchy is then formed by concurrently merging locally
optimal community pairs that increase modularity, repeating this until no
further modularity increases are possible. As all decisions on which 
communities to merge are based on purely local information, a natural 
counterpart strategy exists for local clustering. 

The motivation for \GLO clustering was to improve the computational performance and result quality
of community identification by favoring the formation of a better
hierarchy. The performance improvements have been largely achieved. The
hierarchical structure, as encoded in the dendrogram, is considerably
better balanced than that produced by SSG clustering, with corresponding
improvements in computational performance observed for both model and
empirical networks. The hierarchies produced by \GLO clustering are
moderately worse than those produced by MSG clustering, which is far
more aggressive about making merges.

In terms of the modularity of the community solutions, the best results
are found for the model networks using SSG clustering. But the results
with the model are not borne out in reality---the highest modularity
solution is found with \GLO clustering for fifteen of the twenty empirical
networks considered, including the eight largest networks. 

Overall, the local optimality condition proposed in this paper appears
to be a good basis for forming clusters. We can gain some insight into
this from the local clustering algorithm. For each of the empirical
networks considered here, there is some overlap of the communities,
with several networks showing a great deal of community overlap. The
borders between communities are then not entirely well defined, with
the membership of particular vertices depending on the details of the
sequence of merges performed in partitioning the vertices. The concurrent
building of communities in \GLO clustering seems to allow suitable
cores of communities to form, with the local optimality condition providing a
useful basis for identifying those cores.

Several directions for future work seem promising. First, the local
clustering algorithm described in section\nobreakspace \ref {sec:clusterexpansion} has
worst-case time complexity \( \worstcase{n^3} \) and is thus unsuited to
investigation of large networks; a reconsideration of the local algorithm
may lead to a method suited to a broader class of networks. 
Second,
we observe that nothing about \GLO clustering requires that it be
used with modularity, so it may prove beneficial to apply \GLO clustering 
to community quality measures for specialized classes of networks (\eg, bipartite
networks \citep{Bar:2007}) or to quality measures based on substantially different 
assumptions than modularity (\eg, ratio association \citep{AngBocMarPelStr:2007} or the map equation \citep{RosAxeBer:2009}).  
Finally, we note that
\GLO clustering need not be used with networks at all; application to
broader classes of data analysis could thus be explored, developing \GLO
clustering into a general tool for classifying data into an informative
hierarchy of clusters.

\begin{acknowledgement}
This work has been partly funded by 
the Austrian Science Fund (FWF): [I 886-G11] 
and 
the Multi-Year Research Grant (MYRG) – Level iii (RC Ref. No. MYRG119(Y1-L3)-ICMS12-HYJ) by the University of Macau.
\end{acknowledgement}


\bibliographystyle{unsrtnat}

\bibliography{loc}

\end{document}